\documentclass[aps,pre,floats,floatfix,superscriptaddress,twocolumn,final]{revtex4-2} 
\usepackage{graphicx}
\usepackage{dcolumn,bm,float}
\usepackage{amssymb,amsmath,amsfonts}
\usepackage{adjustbox}
\usepackage{color,url}
\usepackage[usenames,dvipsnames]{xcolor}
\usepackage{ragged2e}
\usepackage{booktabs}
\usepackage[percent]{overpic}
\usepackage{enumitem}
\usepackage{makecell}
\usepackage[hidelinks]{hyperref}
\hypersetup{colorlinks=false,
    linkcolor=black,
    citecolor=black,
    urlcolor=black,
    pdftitle={Mapping biased random walks},
    pdfauthor={Anton Holmgren}
}
\usepackage{xurl}
\usepackage{comment}
\usepackage[capitalise]{cleveref}
\usepackage{orcidlink}
\crefname{section}{Sec.}{Sections}
\crefname{Section}{Sec.}{Sections}

\setcitestyle{numbers,super,comma,sort&compress}

\DeclareMathOperator{\jsd}{JSD}

\newcommand{\subfig}[1]{{\sffamily\bfseries{#1}}}

\makeatletter
\newcommand*{\balancecolsandclearpage}{%
  \close@column@grid
  \cleardoublepage
}
\makeatother

\usepackage{helvet}

\usepackage{booktabs}
\usepackage{tabularx}
\newcolumntype{Y}{>{\raggedleft\arraybackslash}X}

\usepackage{tikz}
\usepackage{siunitx}
\begin{document}

\title{Mapping memory-biased dynamics with compact models reveals overlapping communities in large networks}
\author{Maja Lindstr\"om\,\orcidlink{0009-0009-9224-4646}$^{\ddagger}$}
\email{maja.lindstrom@umu.se}
\affiliation{Department of Computing Science, MIT-huset, Ume{\aa} University, SE-901 87 Ume{\aa}, Sweden}
\affiliation{Integrated Science Lab, Department of Physics, Ume{\aa} University, SE-901 87 Ume{\aa}, Sweden}
\affiliation{Siftlab AB, D{\"o}belnsgatan 12, SE-113 58 Stockholm, Sweden}
\altaffiliation{These authors contributed equally to this work.}

\author{Rohit Sahasrabuddhe\,\orcidlink{0000-0002-2779-8310}$^{\ddagger}$}
\email{rohit.sahasrabuddhe@maths.ox.ac.uk}
\affiliation{Mathematical Institute, University of Oxford, United Kingdom}
\affiliation{Institute for New Economic Thinking, University of Oxford, United Kingdom}
\altaffiliation{These authors contributed equally to this work.}

\author{Anton Holmgren\,\orcidlink{0000-0001-5859-4073}}
\affiliation{Integrated Science Lab, Department of Physics, Ume{\aa} University, SE-901 87 Ume{\aa}, Sweden}

\author{Christopher Bl\"ocker\,\orcidlink{0000-0001-7881-2496}}
\affiliation{Integrated Science Lab, Department of Physics, Ume{\aa} University, SE-901 87 Ume{\aa}, Sweden}
\affiliation{\mbox{Data Analytics Group, Department of Informatics, University of Zurich, CH-8006 Zurich, Switzerland}}
\affiliation{Chair of Machine Learning for Complex Networks, Center for Artificial Intelligence and Data Science (CAIDAS), University of W{\"u}rzburg, DE-97070 W{\"u}rzburg, Germany}

\author{Daniel Edler\,\orcidlink{0000-0001-5420-0591}}
\affiliation{Integrated Science Lab, Department of Physics, Ume{\aa} University, SE-901 87 Ume{\aa}, Sweden}
\affiliation{Gothenburg Global Biodiversity Centre, University of Gothenburg, Gothenburg, Sweden}

\author{Martin Rosvall\,\orcidlink{0000-0002-7181-9940}}
\affiliation{Integrated Science Lab, Department of Physics, Ume{\aa} University, SE-901 87 Ume{\aa}, Sweden}
\affiliation{Siftlab AB, D{\"o}belnsgatan 12, SE-113 58 Stockholm, Sweden}

\begin{abstract}
    Many real-world systems, from social networks to protein-protein interactions and species distributions, exhibit overlapping flow-based communities that reflect their functional organisation.
    However, reliably identifying such overlapping flow-based communities requires higher-order relational data, which are often unavailable.
    To address this challenge, we capitalise on the flow model underpinning the representation-learning algorithm \textsl{node2vec} and model higher-order flows through memory-biased random walks on first-order networks.
    Instead of simulating these walks, we model their higher-order dynamic constraints with compact models and control model complexity with an information-theoretic approach.
    Using the map equation framework, we identify overlapping modules in the resulting higher-order networks.
    Our compact-model approach proves robust across synthetic benchmark networks, reveals interpretable overlapping communities in empirical networks, and scales to large networks.
\end{abstract}

\maketitle

\section{Introduction}

Network-theoretic tools offer a powerful framework for modelling, analysing, and understanding the organisation and function of complex systems~\cite{posfai2016network,newman2018networks}.
These systems often consist of tightly knit modules, or communities, that represent distinct functional parts~\cite{fortunato2016community}.
In many networks, these communities overlap: nodes may belong to more than one community, with overlaps ranging from rare exceptions to extensive intersections~\cite{xie2013overlapping}.

Popular methods for detecting overlapping communities rely on topological heuristics, such as clique percolation~\cite{palla2005uncovering}, local optimisation of order statistics~\cite{lancichinetti2011finding}, and link communities~\cite{ahn2010link}, recently extended to  hypergraphs~\cite{lotito2024hyperlink}.
But many real-world networks, particularly those with weighted or directed links, represent more than local structure; they impose constraints on network flows that connect non-adjacent nodes~\cite{rosvall2008maps,scholtes2014causality,masuda2017random,ballal2022network,bovet2022flow}.

Mapping such flows naturally reveals overlapping communities. For example,  
citation paths traverse through overlapping research fields in scholarly publishing~\cite{rosvall2014memory}, global maritime traffic move through overlapping clusters of international ports~\cite{xu2016representing}, and information circulate within overlapping groups of friends and colleagues in social networks~\cite{sahasrabuddhe2025concise}.
These flow patterns define the system's organisation: groups of nodes that trap flows form overlapping flow-based communities.
When higher-order relational data are available, such overlapping communities can be identified directly without relying on topological heuristics~\cite{rosvall2014memory,edler2017mapping,xu2016representing,lambiotte2019networks,sahasrabuddhe2025concise}.

Without higher-order network data, flow-based methods impose hard boundaries or minimal overlap~\cite{esquivel2011compression}, misrepresenting the system's organisation.
For example, applying a memoryless Markov model to a journal citation network assigns a multidisciplinary journal like Nature to a single research field, neglecting its broader role~\cite{persson2016maps}.
Since higher-order data are often unavailable, this limitation raises a critical question:
How can we identify overlapping flow-based communities from link data alone?

\begin{figure*}[hpt!]
  \centering
  \includegraphics[width=0.88\textwidth]{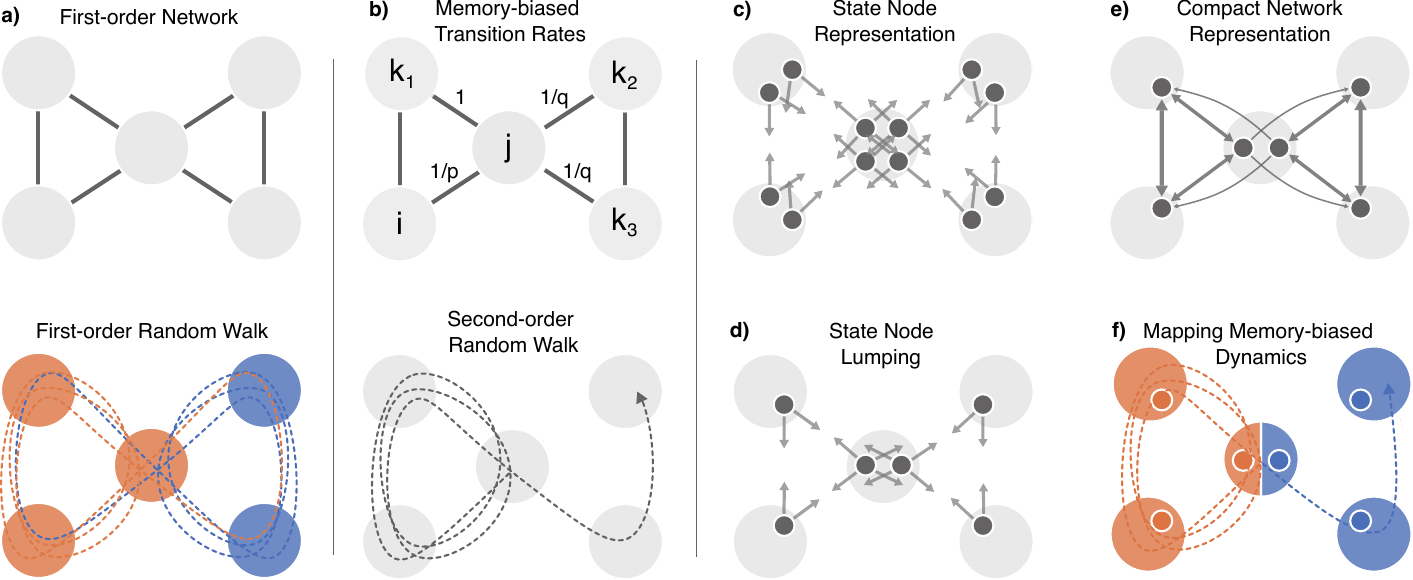}
  \caption{\textbf{A compact representation of higher-order dynamics reveals overlapping communities.} 
  \subfig{a)}~A first-order network constrains a memoryless random walk that supports two non-overlapping communities. \subfig{b)}~We introduce a second-order model by biasing transitions: arriving along the link $(i, j)$, the random walker at node $j$ is biased to backtrack by a factor $1/p$ and to move to a node not adjacent to $i$ by a factor $1/q$.
  \subfig{c)}~We describe the bias in each physical node with state nodes, where arrows indicate transition probabilities. \textbf{d)} To manage computational complexity, we lump state nodes within each physical node.
  \subfig{e)}~By connecting the lumped state nodes, we construct a compact network representation. 
  \subfig{f)}~Mapping the memory-biased random walk on the compact network reveals overlapping communities in the middle node.}
  \label{fig:schematic}
\end{figure*}
To reveal overlapping communities, we approximate higher-order dynamics on first-order networks using memory-biased random walks.
In the spirit of memory models with tuned rates for return, triangular, and exploratory steps~\cite{rosvall2014memory},
we focus on a second-order model inspired by the representation learning algorithm \textsl{node2vec}~\cite{grover2016node2vec}.
A random walker at node~$j$ remembers the previously visited node~$i$ and selects the next node~$k$ based on the shortest-path distance between~$i$ and~$k$~(\cref{fig:schematic}). By weighting transitions by distance, we can adjust the walker's exploration radius to favour local or global flow patterns.

Bypassing explicit random walk simulations, we model their constraints using memory networks~\cite{edler2017mapping}, where \textit{physical nodes} mirror the original nodes in the first-order network and \textit{state nodes} track transitions that depend on the previous step.
To manage the combinatorial explosion of second-order paths, we build a compact representation~\cite{persson2016maps, sahasrabuddhe2025concise} that retains essential flow dynamics while cutting model complexity.
We then apply the map equation framework~\cite{rosvall2008maps,rosvall2009map,mapequation2022software} on the compact higher-order networks to detect overlapping communities~(\cref{fig:schematic}).

The proposed method proves robust across synthetic benchmarks and effectively identifies interpretable overlapping flow-based communities in real-world networks.
The compact higher-order models scale to large networks while retaining the prominent flow constraints of memory-biased random walks.

\section{Modelling memory-biased random walks}
The simplest model of flow on a network captures movement constrained solely by the out-links. In an unbiased first-order random walk, a walker at node $j$ transitions to node $k$ with rate
\begin{equation}
    \pi_{jk} = \frac{w_{jk}}{\sum_{k'} w_{jk'}},
\end{equation}
where $w_{jk}$ is the weight of the link $j \rightarrow k$. The transitions depend only on the current position in this memoryless model.

To model memory effects, second-order random walks incorporate the previous step. The transition from $j$ to $k$ depends not only on $j$ but also on the node $i$ from which the walker arrived. The \textsl{node2vec} walk~\cite{grover2016node2vec} captures such memory by biasing the first-order transition rate depending on whether $i$ and $k$ are neighbours,
\begin{equation}
     B(i,k) = 
         \begin{cases}
            1 / p \quad \text{if } i = k,\\
            1 \qquad \text{if $i$ and $k$ are neighbours},\\
            1 / q \quad \text{otherwise}.
        \end{cases}
\end{equation}
The parameters $p$ and $q$ control the rate of diffusion: $p$ influences the likelihood of going back to the previous node, while $q$ affects the tendency to move to a node not neighbouring $i$.

\subsection{Full second-order representation}
To fully capture the \textsl{node2vec} dynamics, we construct the weighted and directed second-order network $M_2$. For every physical node $j$ with in-degree $d^\text{in}_j$, we create state nodes $j_i$ for each in-link $i \rightarrow j$, representing arrivals at $j$ from $i$. The link weight between state nodes $j_i$ and $k_j$ is the physical link weight $w_{jk}$ divided between the state nodes $j_i$ to preserve the physical network and multiplied by the biased transition rates

\begin{equation}
    w_{j_i k_j} = \frac{w_{jk}}{d^\text{in}_j}
    B(i,k).
\end{equation}

For physical nodes without in-links, we create a single state node and use the first-order transition rates. With this state node network representation, the map equation framework~\cite{edler2017mapping} can reveal modular structures in the higher-order dynamics. To compress the network dynamics, Infomap can split a physical node's state nodes across overlapping modules.

The second-order model of a physical node $j$ with in- and out-degrees $d_j^\text{in}$ and $d_j^\text{out}$, has $d_j^\text{in}$ state nodes, each with $d_j^\text{out}$ out-links. For a network with $n$ nodes and $m$ links, $M_2$ has $m$ nodes, or $2m$ if undirected, and $\sum_{j=1}^n d_j^\text{in} d_j^\text{out}$ links, making analysis computationally demanding for large networks.

\subsection{Compact second-order representation}
To efficiently analyse large networks, we lump state nodes into a compact second-order model that preserves key flow dynamics. For a partition $\mathcal{P}^j$ of physical node $j$'s state nodes, we create lumped state nodes $j_\alpha$ for each subset $\alpha \in \mathcal{P}^j$. Each lumped node $j_\alpha$ represents arrivals at $j$ from any node $i \in \alpha$. To preserve flow dynamics when aggregating links, we weight them according to arrival frequencies. The link weight between lumped nodes $j_\alpha$ and $k_\beta$ is

\begin{equation}
    w_{j_\alpha k_\beta} = 
    \sum_{i \in \alpha} \frac{p_{i_j}}{p_\alpha} d_j^\text{in} w_{j_i k_j},
    \label{eq:lumped_weight}
\end{equation}
where $p_{j_i}$ is the PageRank flow of state node $j_i$, and $p_\alpha = \sum_{i \in \alpha} p_i$ is the aggregated flow for the lumped node $j_\alpha$.

To measure the information loss by replacing $M_2$ with the compact model $M_c$, we use the Jensen--Shannon divergence with the Shannon entropy $H$ measured in bits. For physical node $j$, the Jensen--Shannon divergence is
\begin{equation}
    \jsd(M_c^j, M_2^j) = \overbrace{\sum_{\alpha \in \mathcal{P}^j} p_\alpha H(\bm{\pi_{j_\alpha}})}^{M_c^j} - \overbrace{\sum_{i} p_{j_i} H(\bm{\pi_{j_i}})}^{M_2^j},
    \label{eq:jsd_mv_m2}
\end{equation}
where $\bm{\pi_{j_i}} = w_{j_i \cdot} / \sum_{k'} w_{j_i k'_j} $ is the transition rates from state node $j_i$ and 
\begin{equation}
    \bm{\pi_{j_\alpha}} = \sum_{i \in \alpha} 
    \frac{p_{i_j}}{p_\alpha} \bm{\pi_{j_i}}
\end{equation}
the flow-weighted lumped transition rates.

For the entire network,
\begin{equation}
    \jsd(M_c, M_2) = \sum_j\jsd(M_c^j, M_2^j),
    \label{eq:jsd_entire_network}
\end{equation}
which is the difference in entropy rates between the compact and $M_2$ model.
When the random walk is unbiased $(p=q=1)$, all transition-rate vectors $\bm{\pi_{j_i}}$ at each physical node coincide so that $\jsd(M_c, M_2)=0$.

We interpolate between a first-order model $M_1$ and $M_2$ by iteratively dividing the compact node with the highest $\jsd(M_c, M_2)$ in two using a divisive clustering algorithm.

\begin{figure}[hpt!]
    \centering
    \includegraphics[width=\linewidth]{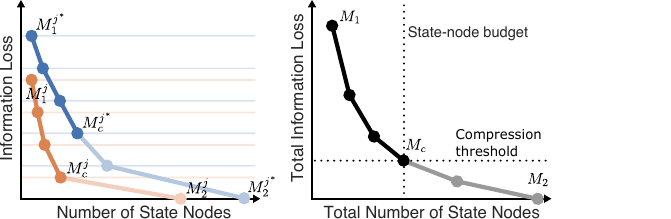}
    \caption{\textbf{Compact model information loss.} \textbf{Left:}~Per-physical-node information loss as a function of the number of state nodes. Trajectories shown for two physical nodes, with $j^*$ denoting the physical node with the largest information loss. \textbf{Right:}~The corresponding total information loss summed over all physical nodes. In this schematic example, the fourth compact model $M_c$ represents the model either constrained by the compression threshold or the user-defined state-node budget.}
    \label{fig:entropy}
\end{figure}

\subsubsection{Divisive clustering algorithm}
\label{sec:divisive}

The divisive clustering algorithm splits lumped state nodes in each physical node to minimise information loss.
Given parameters $p$ and $q$, we assemble the compact network model by
\begin{enumerate}[label={(\roman*)},itemsep=-2pt]
    \item creating lumped state nodes  independently for each physical node, and
    \item adding links between lumped state nodes $j_\alpha \rightarrow k_\beta$ for $j \in \beta$ with weight $w_{j_\alpha k_j}$ (Eq.~\ref{eq:lumped_weight}).
\end{enumerate}

\begin{figure}[hpt!]
    \centering
    \begin{overpic}[width=0.8\linewidth]{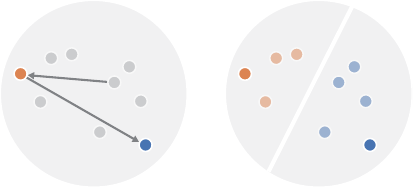}
        \put(25, 28){$i_0$}
        \put(3, 30){$i_1$}
        \put(33, 13.5){$i_2$}
    \end{overpic}
    \caption{\textbf{The divisive clustering algorithm.}
    Schematic embedding of state nodes in cluster $\alpha$, where each point denotes a state node and distances reflect the Jensen--Shannon divergence between their transition-rate vectors.
    \textbf{Left:}~To cluster the state nodes, we first pick a random state node $i_0$. Then, we select the cluster centres $i_1$ as the state node farthest from $i_0$ and $i_2$ as the state node farthest from $i_1$.
    \textbf{Right:}~We assign the state nodes to the nearest cluster centre.}
    \label{fig:lumping}
\end{figure}
We propose a fast divisive clustering algorithm that partitions state nodes in each physical node to minimise information loss. From Eq.~\ref{eq:jsd_mv_m2}, the Jensen--Shannon divergence of cluster $\alpha$ of physical node $j$ is
\begin{equation}
    p_\alpha H(\bm{\pi_{j_\alpha}}) - \sum_{i \in \alpha} p_i H(\bm{\pi_{j_i}}).
\end{equation}

Starting from the first-order model, a single cluster containing all state nodes of a physical node, we iteratively split the cluster with the highest divergence into two clusters with similar state nodes. We continue until the total compression $\Delta_\text{JSD} = 1 - \jsd(M_c, M_2) / \jsd(M_1, M_2)$ exceeds a compression threshold $\Delta_\text{MIN}$, default 90 percent, or we reach an optional state node budget, a limit on the number of state nodes. To split cluster $\alpha$, we:
\begin{enumerate}[label={(\roman*)},itemsep=-2pt]
    \item pick a random seed state node $i_0 \in \alpha$;
    \item identify two cluster centres by Jensen--Shannon distance, $i_1 \in \alpha$ farthest away from $i_0$ and $i_2 \in \alpha$ farthest away from $i_1$;
    \item assign each state node in $\alpha$ to its nearest centre (Fig.~\ref{fig:lumping}).
\end{enumerate}
Because this procedure is fast and stochastic, we repeat these steps several times and retain the division with the lowest divergence.

Each iteration scales linearly with cluster size. A physical node with degree $d$ has $d$ state nodes. Identifying cluster centres and assigning state nodes requires $\mathcal{O}(d)$ operations. Splitting each node at most $d$ times across all $n$ nodes gives $\mathcal{O}(nd^2) = \mathcal{O}(md)$ complexity. The algorithm thus scales linearly with the number of links for bounded average degree.

\section{Mapping memory-biased random walks}

To detect communities, we apply Infomap~\cite{mapequation2022software}, which searches for a network partition $\mathsf{M}$ that minimises the expected codelength $L(\mathsf{M})$ for a modular description of a random walk on the network~\cite{rosvall2008maps}.
For each network and set of parameters, we run 20 independent Infomap optimisation trials and select the one with the lowest codelength.
For clarity in illustrations, we constrain the search to two-level partitions, though the proposed approach naturally extends to hierarchical partitions.

Our evaluation strategy reflects a conceptual difference in how we represent overlapping communities. Most overlapping community-detection algorithms assign nodes to communities with fuzzy membership weights, producing a single partition of the node set with probabilistic assignments. Our state-node approach produces a different object: it identifies which flow paths through a node participate in which communities. A node visited from different predecessors can play distinct roles in different communities. This representation captures constraints on higher-order flows that fuzzy membership cannot express. We therefore focus our evaluation on internal validation. First, how well do compact models approximate complete higher-order dynamics? Second, do detected communities reflect interpretable structures in real-world networks? Comparisons with established methods serve to illustrate sensible behaviour, not to rank performance on standardised metrics.

\subsection{Synthetic networks}
We test our method on a schematic network to illustrate detected overlapping communities and on benchmark networks to evaluate how well compact models approximate full second-order models.

\subsubsection{Schematic network}
\begin{figure}[htp!]
    \centering
    \vspace*{2cm}
    \begin{overpic}[width=1\linewidth]{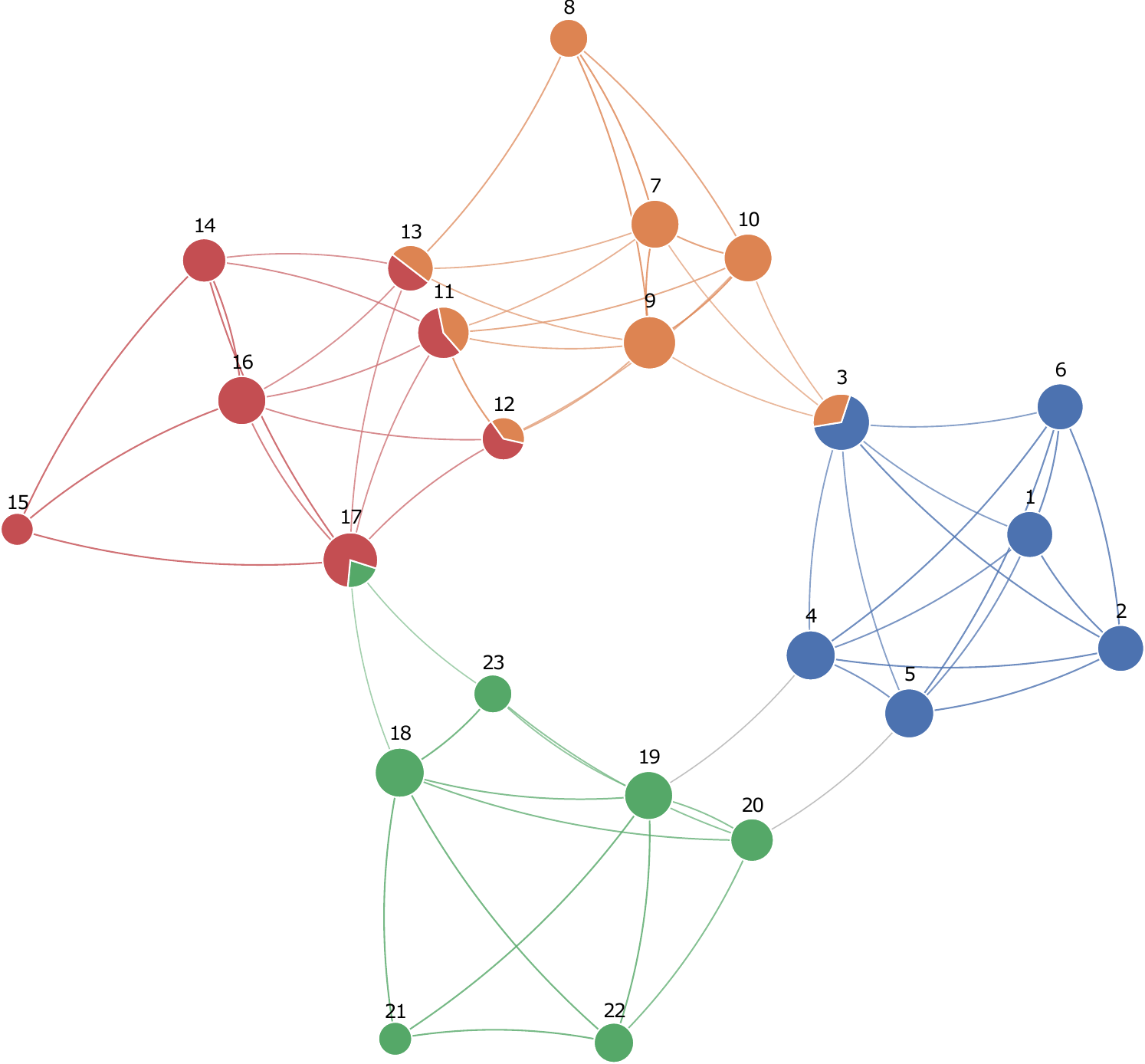}
        \put(7,110){\subfig{a)}}
        \put(7,81){\subfig{b)}}
        \put(10,84){\includegraphics[width=0.35\linewidth]{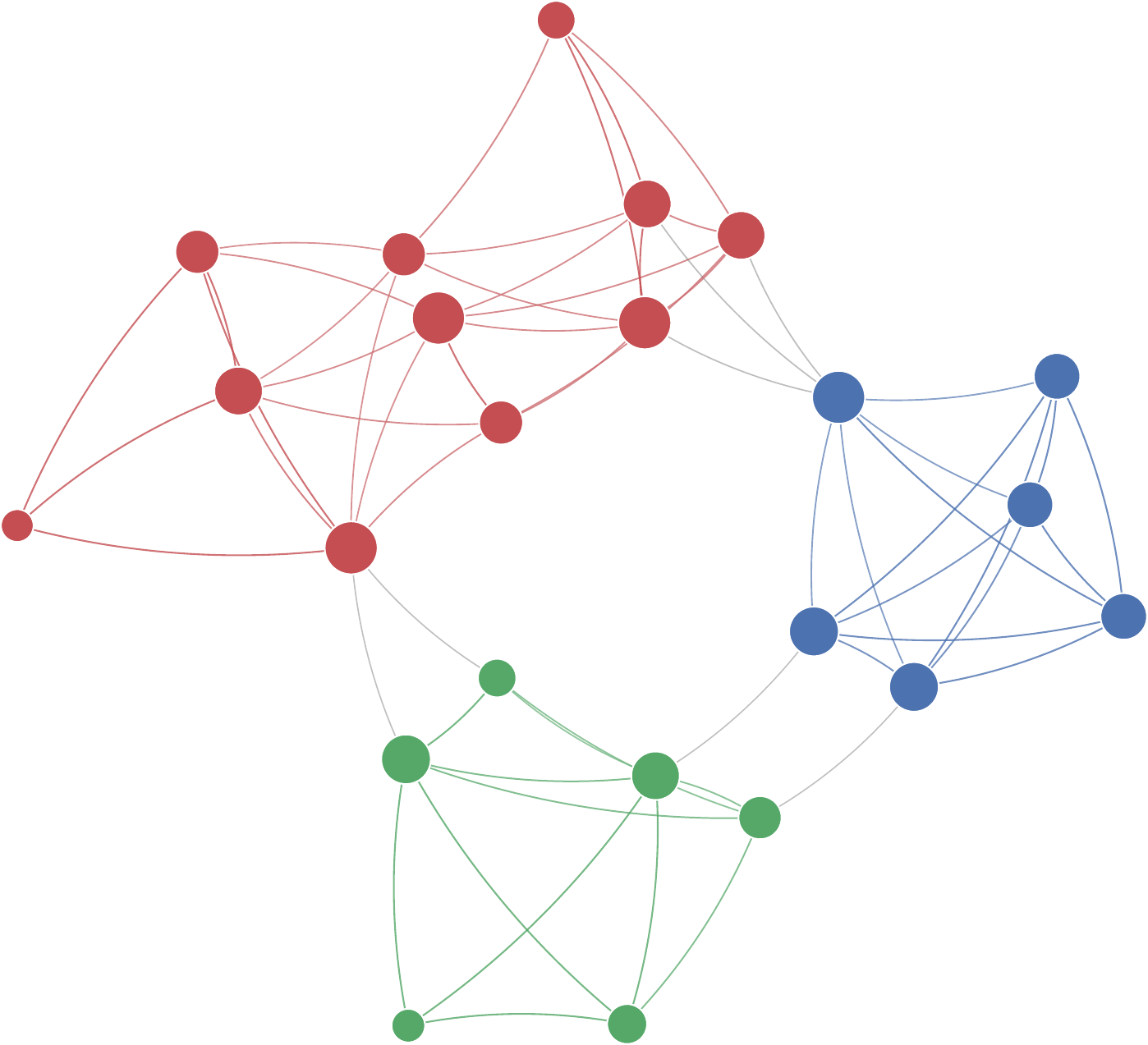}}    
    \end{overpic}
    \caption{\textbf{Synthetic network with overlapping community structure.}
    \subfig{a)}~The first-order network with three modules.
    \subfig{b)}~With $p=1$, $q=2$, we find four modules and five nodes that are in multiple communities.}
    \label{fig:example}
\end{figure}

To showcase the full second-order model, we apply it to the hand-crafted schematic network designed by Palla et al.\cite{palla2005uncovering} to illustrate overlapping communities (\cref{fig:example}). We set $p=1$ and $q=2$ and identify the previously reported overlapping communities and, unlike the original study, also detect a small overlap at node~17.

\subsubsection{LFR benchmark networks}
To assess the accuracy, we construct instances of the LFR benchmark networks with planted overlapping communities~\cite{lancichinetti2009benchmarks} using $N = 100$ nodes, average node degree $k = 6$, clustering coefficient $C = 0.3$, min module size $c_{\min} = 10$, max module size $c_{\max} = 20$, max degree $k_{\max} = 15$, and mixing parameter $\mu = 0.2$.

We vary the number of overlapping nodes $o_n \in \{5, 15\}$, and the number of memberships of the overlapping nodes $o_m \in \{2, 4\}$. To assess the effect to the bias parameters, we use $p \in \{0.5, 1\}$ and $q \in \{2, 4\}$. To compute average performance, we generate 20 network instances for each parameter combination. For each network instance, we create 10 compact models $M_c$ with different state node budgets. We run Infomap 20 times on each compact model to find modules that retain the memory-biased flow, which may differ from those planted in the LFR framework.

\begin{figure}[hpt!]
    \centering
    \includegraphics[width=\linewidth]{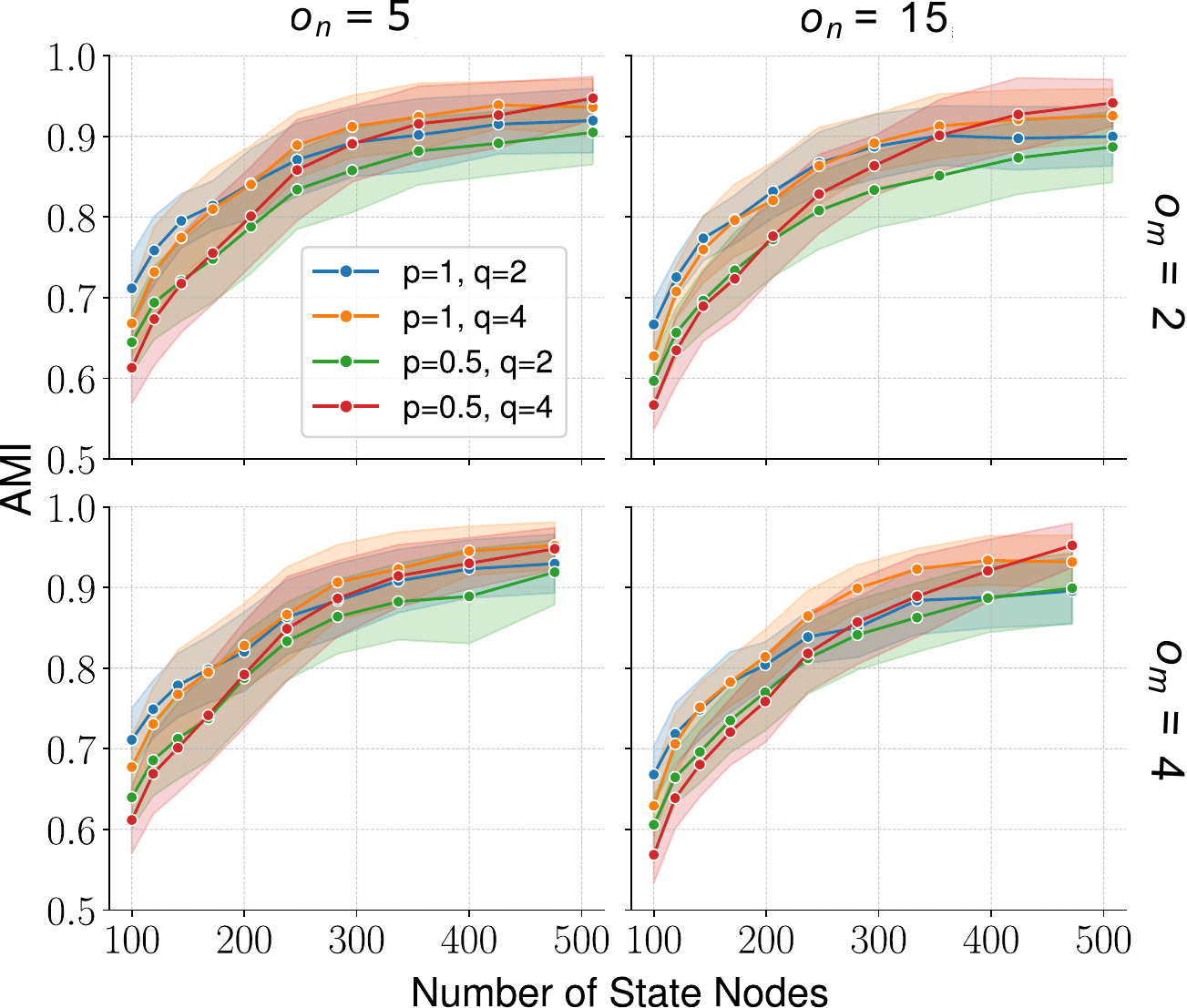}
    \caption{\textbf{Accuracy on LFR networks.}
    The AMI between the community structures of the compact models and the full second-order models as a function of the state-node budget.
    We plot the mean AMI across 20 network instances, with shaded regions showing one standard deviation.}
    \label{fig:benchmark}
\end{figure}

To measure the accuracy, we compare the community structure of $M_c$ to that of the full second-order model $M_2$. Across different values of $p$, $q$, $o_n$, and $o_m$, the Adjusted Mutual Information (AMI) remains high (\cref{fig:benchmark}). Because we compare with one particular $M_2$ solution, we do not expect the AMI to reach its maximum value of one. The solution landscape is flat: different partitions can have similar codelengths yet look quite different~\cite{calatayud2019exploring}. When we evaluate compact models with few state nodes, we compare their AMI against that of larger models. As the number of state nodes increases, the AMI rises sharply at first, showing that even relatively small compact models capture important effects of the bias. When $M_c$ contains about half as many state nodes as $M_2$, the AMI stabilises near $0.9$, demonstrating that we can closely match the community structure of the full second-order network with a much smaller model.

\subsubsection{ABCD$+o^2$ benchmark networks}
To extend our analysis on synthetic benchmark networks, we include the Artificial Benchmark for Community Detection with Outliers and Overlapping Communities (ABCD$+o^2$) \cite{ kaminski2021artificial, barrett2025artificial}. This benchmark produces large and realistic networks characterised by heavy-tailed degree and community-size distributions, with scalability to millions of nodes. We generate ABCD$+o^2$ instances with $N = 100$ nodes, $N_{\text{out}} = 1$ outlier, power-law exponent for the degree distribution $t_1 = 2$, minimum degree $d_{\min} = 6$, maximum degree $d_{\max} = 15$, power-law exponent for the community-size distribution $t_2 = 1$, minimum community size $c_{\min} = 10$, maximum community size $c_{\max} = 20$, fraction of edges in the background graph $\xi = 0.2$,  latent space dimensionality $d = 2$, and correlation between node degree and number of communities $\rho = 0.0$.

\begin{figure}[hpt!]
    \centering
    \includegraphics[width=\linewidth]{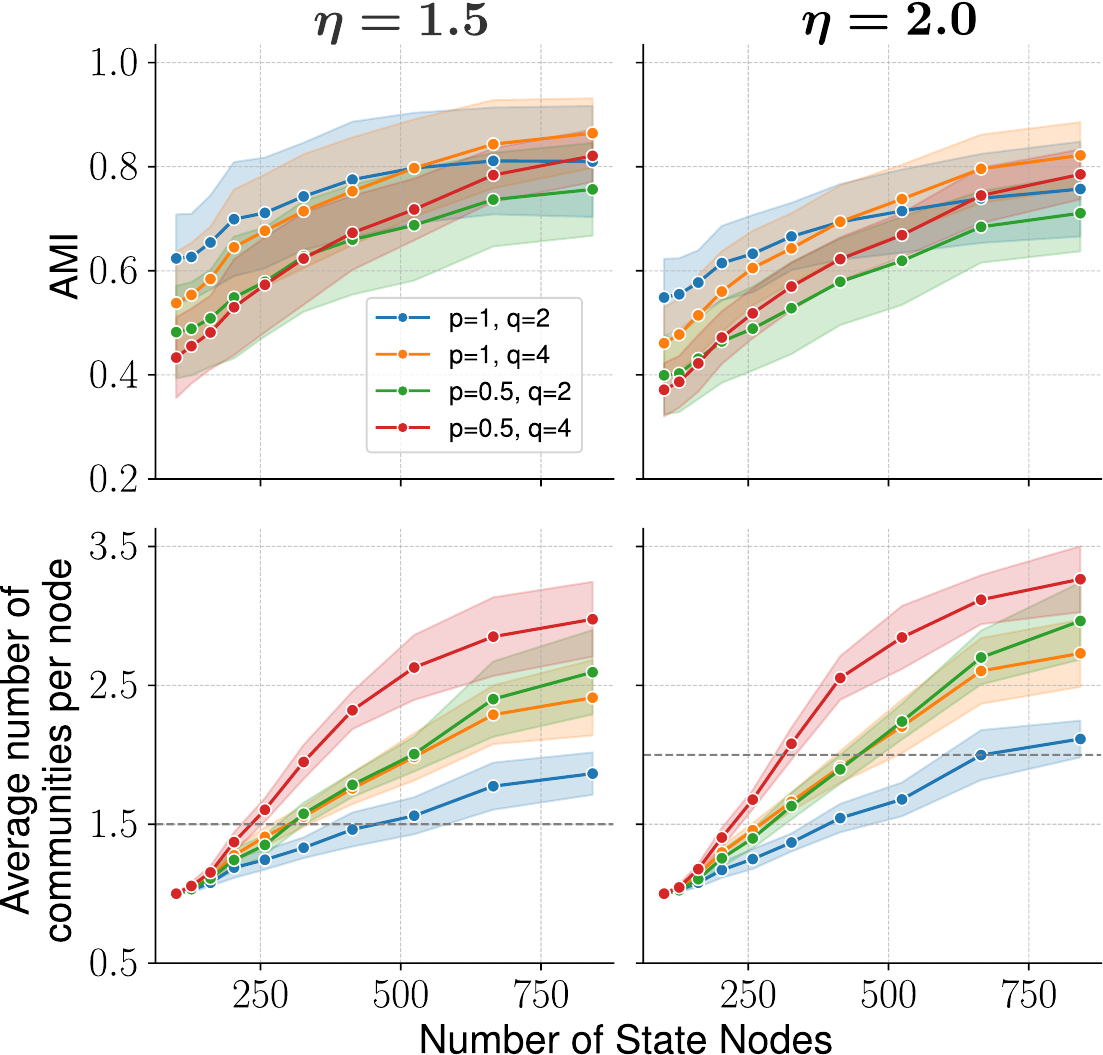}
    \caption{\textbf{Accuracy on ABCD$\bm{+o^2}$ networks.}
    Top row shows the AMI between the community structures of the compact models and the full second-order models as a function of the state-node budget. Bottom row shows the average number of communities per node for the compact models. Dashed lines indicate the $\eta$ values, the planted average number of communities per node. We plot the mean across 20 network instances, with shaded regions showing one standard deviation.}
    \label{fig:benchmark_abcd}
\end{figure}

We evaluate these networks as the LFR benchmarks, but instead of varying $o_n$ and $o_m$, we vary the average number of communities per node, $\eta \in \{1.5, 2.0\}$. We measure both the AMI between compact and full second-order models and the average number of communities per node in the detected overlapping communities (\cref{fig:benchmark_abcd}).

As with the LFR networks, the AMI first increases quickly with the number of state nodes. Capturing full second-order models with more overlap requires more state nodes: models with low $p$ and high $q$ produce higher overlap and achieve lower AMI with the same state-node budget. The compact models successfully identify more overlap when planted networks contain more overlap. Models with $p=1$ and $q=2$, which bias walks toward neighbours of the source node, best capture the planted structures.

\subsection{Real-world networks}

We apply our method to real-world networks to assess whether it reveals overlapping communities that reflect explainable structure. Specifically, we analyse a word association network centred on the word \emph{bright}, a network of teams competing in U.S.\ college football, and a character association network from a fantasy novel (\cref{tab:dataset_summary}). We then compare our compact memory-biased model with established overlapping community detection methods and with a full second-order memory model.

\subsubsection{Word association network}
\begin{figure}
    \centering
    \includegraphics[width=1\linewidth]{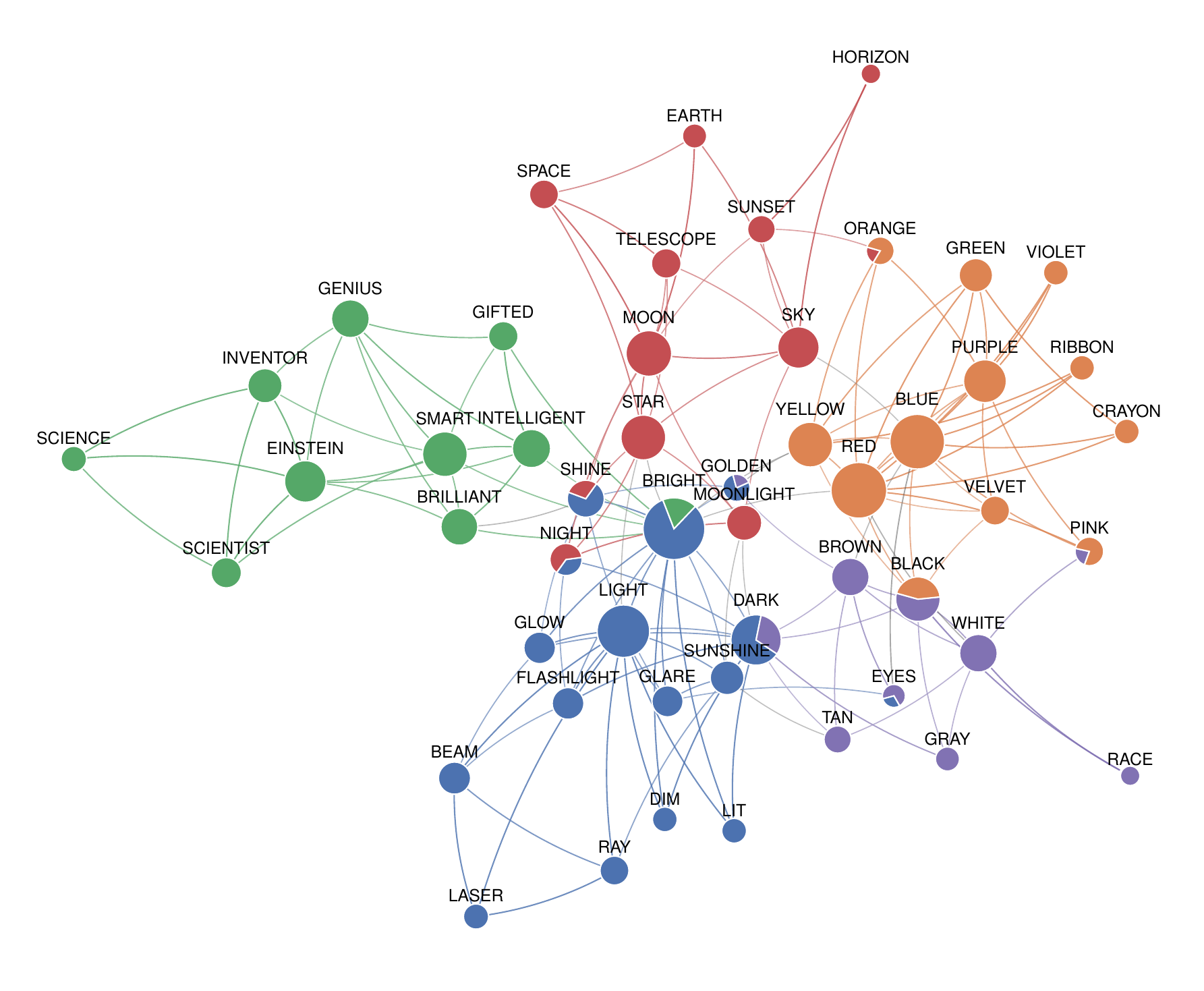}
    \caption{\textbf{Overlapping communities in the word association network.} Node sizes are proportional to their flow volume. When a node is in multiple modules, we show a pie chart where the slices indicate the distribution of flow. Link widths and saturation are proportional to the flow between nodes. The compact model uses about $50$ percent of the state nodes, with $p=1$ and $q=2$.}
    \label{fig:word_assoc}
\end{figure}
The word association network represents free associations between words collected by researchers at the University of South Florida\footnote{\url{http://w3.usf.edu/FreeAssociation/Intro.html}}. We use the network centred around the word \emph{bright}, as studied in Ref.~\citenum{palla2005uncovering}, where undirected edges connect words which are associated with each other above a given frequency threshold.

We generate a compact network model using approximately $50$ percent of the state nodes with $p=1$ and $q=2$. Infomap identifies five communities, each corresponding to a distinct concept: \emph{intelligence} in green, \emph{light} in blue, \emph{appearance} in purple, \emph{colours} in orange, and \emph{astronomy} in red. The modules overlap at several words with multiple meanings. For instance, both \emph{bright} and \emph{dark} belong to the \emph{light} module, while \emph{bright} also appears in the \emph{intelligence} module and \emph{dark} in the \emph{appearance} module. Similarly, while \emph{moon} and \emph{star} are found only in the \textit{astronomy} module, \textit{night} also appears in the \textit{light} module.

\subsubsection{American college football network}
The American college football network represents games played between Division IA colleges during the fall 2000 regular season. Nodes represent teams, and edges denote games played~\cite{girvan2002community, evans2010clique}. The network exhibits a strong community structure: teams are grouped into conferences and play more games within than outside their conference. In 2000, teams played an average of seven intra-conference games and four inter-conference games. Inter-conference games were unevenly distributed, geographically closer teams were more likely to face each other.

\begin{figure*}[htp!]
    \centering
    \includegraphics[width=0.8\textwidth]{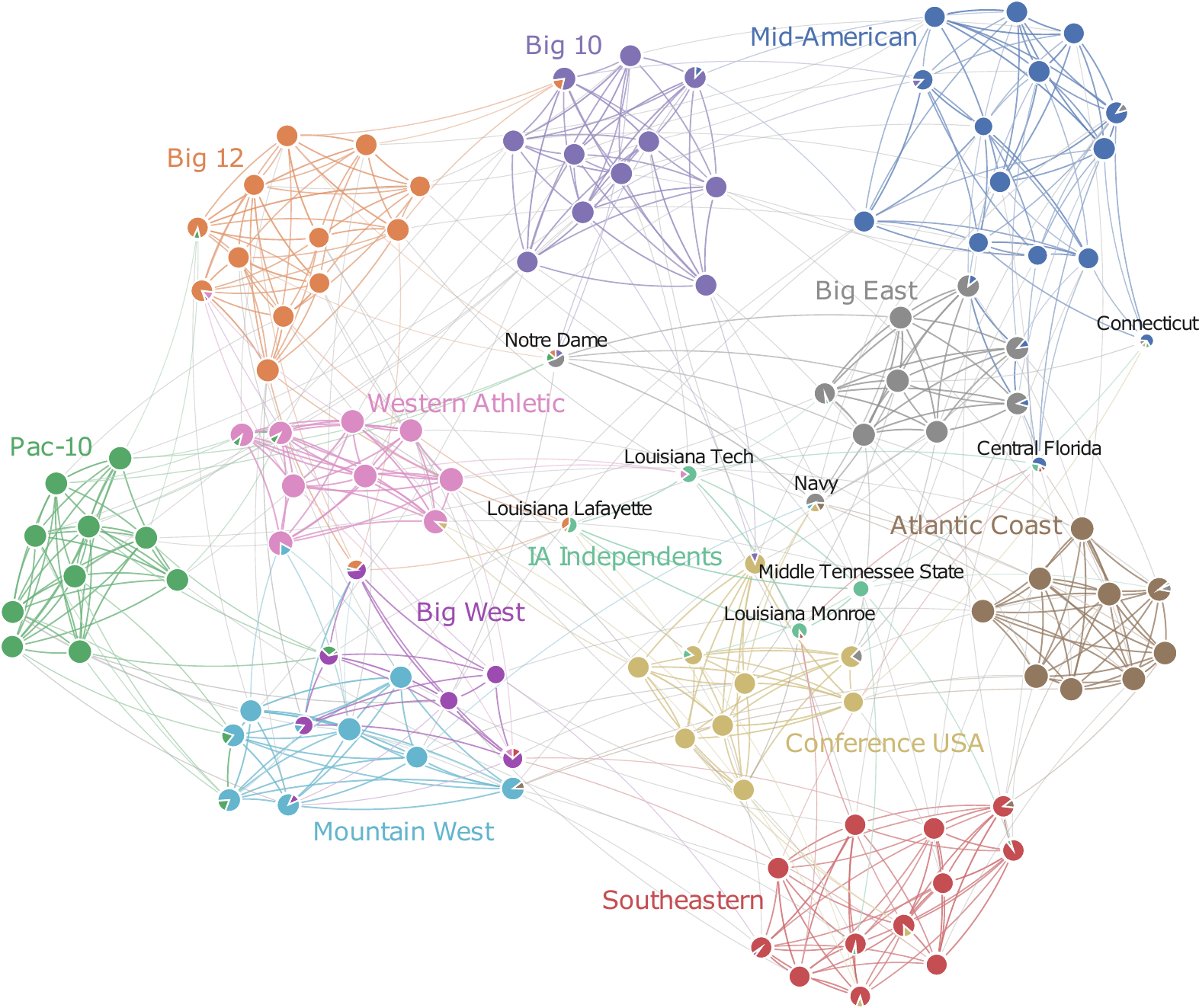}
    \caption{\textbf{American College Football network with overlapping community structure.} Nodes represent teams, and edges denote games played. We observe 12 communities, each labelled with the name of its corresponding conference. Independent teams are shown with names in black. The compact model uses just under $70$ percent of the state nodes, with $p=1$ and $q=4$.}
    \label{fig:football}
\end{figure*}

The first-order solution recovers the known conferences and consolidates nearly all independent teams into a single module, except for \emph{Navy} and \emph{Notre Dame}, which join the \emph{Big East} conference and \emph{Connecticut} which joins the \emph{Mid-American} conference (\cref{table:football}).
Partitioning the compact model, constructed with $p=1$, $q=4$, and just under $70$ percent of the state nodes, reveals $12$ communities. The communities mostly overlap at the independent teams \emph{Central Florida}, \emph{Connecticut}, \emph{Navy}, \emph{Notre Dame}, \emph{Middle Tennessee State}, \emph{Louisiana Tech}, and \emph{Louisiana Monroe} (\cref{table:football}). Additional overlaps arise when teams frequently play multiple inter-conference games against geographically nearby conferences (\cref{fig:football}).

\begin{table*}[hbtp]
\caption{\label{table:football}\textbf{Conference assignment of independent college football teams.} In the first‐order memoryless model, each team is forced into exactly one community, Big East, Mid-American, or its own independent community. In contrast, in the compact memory‐biased model, teams split their membership across multiple conferences, with bar‐length indicating flow percentages.}
\centering
\begin{tabularx}{\textwidth}{@{}r*{8}{Y}@{}}
\toprule
  & \makecell[r]{Central\\Florida}
& \makecell[r]{Connecticut}
& \makecell[r]{Louisiana\\Lafayette}
& \makecell[r]{Louisiana\\Monroe}
& \makecell[r]{Louisiana\\Tech}
& \makecell[r]{Middle\\Tennessee\\State}
& \makecell[r]{Navy}
& \makecell[r]{Notre\\Dame} \\
\midrule
\textbf{First-order model} &  &  &  &  &  &  &  & \\
\addlinespace
Big East & - & - & - & - & - & - & 100 & 100 \\
IA Independents & 100 & - & 100 & 100 & 100 & 100 & - & - \\
Mid-American & - & 100 & - & - & - & - & - & - \\
\midrule
\addlinespace
\textbf{Compact model} &  &  &  &  &  &  &  & \\
\addlinespace
Atlantic Coast & 10 & - & - & - & - & - & 16 & - \\
Big 10 & - & - & - & - & - & - & - & 16 \\
Big 12 & - & - & 36 & - & - & - & - & 14 \\
Big East & - & 12 & - & - & - & - & 56 & 45 \\
Conference USA & - & 11 & 9 & - & - & - & 19 & - \\
IA Independents & 26 & 12 & 55 & 65 & 80 & 79 & - & - \\
Mid-American & 54 & 65 & - & - & - & - & - & - \\
Mountain West & - & - & - & - & - & - & 9 & 8 \\
Pac-10 & - & - & - & - & - & - & - & 17 \\
Southeastern & 10 & - & - & 35 & - & 21 & - & - \\
Western Athletic & - & - & - & - & 20 & - & - & - \\
\end{tabularx}
\end{table*}

\subsubsection{A Storm of Swords network}
We explore overlapping story arcs in the fantasy novel \textit{A Storm of Swords} using its character co-occurrence network~\cite{asoiaf}. The network consists of 303 nodes representing characters and 1,008 edges, where weights count how often their names appear within 15 words of each other. We analyse the community structure of a compact model with 537 state nodes constructed with $p=1$ and $q=4$. Infomap identifies 31 modules, of which 10 include at least 10 characters and together account for 90 percent of the total flow. These modules largely correspond to story arcs in distinct locations such as \textit{The Wall}, \textit{King's Landing}, and \textit{Essos} (\cref{appendix:storm_of_swords}).

Of the 9 central characters with more than 2 percent of the flow each, 6 appear in multiple modules, indicating that they play roles in multiple storylines (Table \ref{tab:storm_of_swords_top15}). For instance, Jon Snow appears in \textit{North of the Wall}, \textit{The Wall}, \textit{Jon's Watch}, and \textit{Winterfell} modules, and Arya Stark is involved in \textit{Arya's adventures}, \textit{The Brotherhood without Banners}, \textit{The Riverlands}, and \textit{Winterfell}.

\subsubsection{Amazon network}

\begin{figure}[htp!]
    \centering
    \includegraphics[width=1\linewidth]{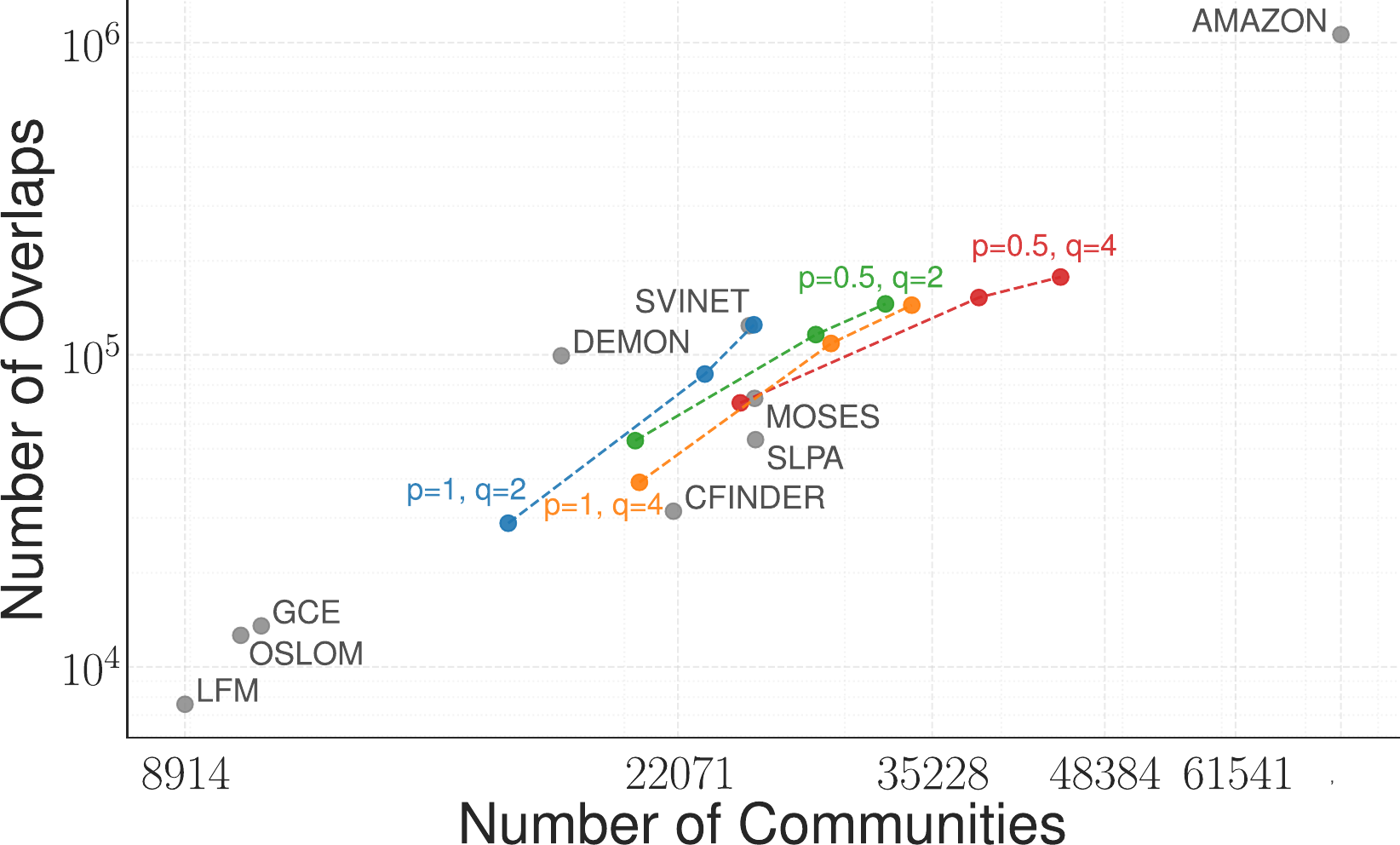}
    \caption{\textbf{Modular overlap in the Amazon network for different methods.} Each point represents a detected community structure, plotted by the number of communities versus the number of overlaps measured as links in the community graph. Three compact memory-biased models with varying $p$ and $q$ values for three different state-node budgets compared with established overlapping community-detection methods. The state-node budgets are 500,000, 750,000, and 1 million state nodes, increasing from left to right. For comparison, the full second-order model comprises 1,851,744 state nodes.}
    \label{fig:amazon}
\end{figure}

To compare our compact memory-biased model with established overlapping community detection methods, we analyse the Amazon product co-purchasing network \cite{amazon}. In this network, nodes represent products and edges connect items frequently bought together. Amazon assigns each product to one or more categories, but this classification serves commercial purposes rather than capturing network structure. The categories produce far more communities and overlaps than any detection method recovers—including LFM \cite{lancichinetti2009detecting}, OSLOM \cite{lancichinetti2011finding}, and our approach (Fig.~\ref{fig:amazon}). Following the work of Jebabli et al.~\cite{jebabli2018community}, we construct a \textit{community graph} where nodes represent communities and edges represent overlaps. To align with their results, we remove singleton nodes from the final community graph.

For the compact memory-biased model, the parameters control overlap in intuitive ways. Lower $p$ values and higher $q$ values increase the tendency to backtrack, causing random walks to revisit physical nodes assigned to multiple communities before settling. This behaviour produces more overlap. The compact models with larger state-node budgets identify more communities with greater overlap than existing methods, though still far fewer than Amazon's category system suggests.

\subsubsection{Taxi drives}

We compare the compact memory-biased model to a model constructed from the complete second-order representation of GPS-tracked taxi drives in San Francisco \cite{rosvall2014memory}. Examining the actual taxi trajectories reveals asymmetric memory: taxis return to recently visited locations less frequently after two steps but more frequently after three steps compared with memoryless expectations. This pattern suggests parameters with $p > 1$ and $q > 1$.

\begin{figure}[htp!]
    \centering
    \includegraphics[width=0.85\linewidth]{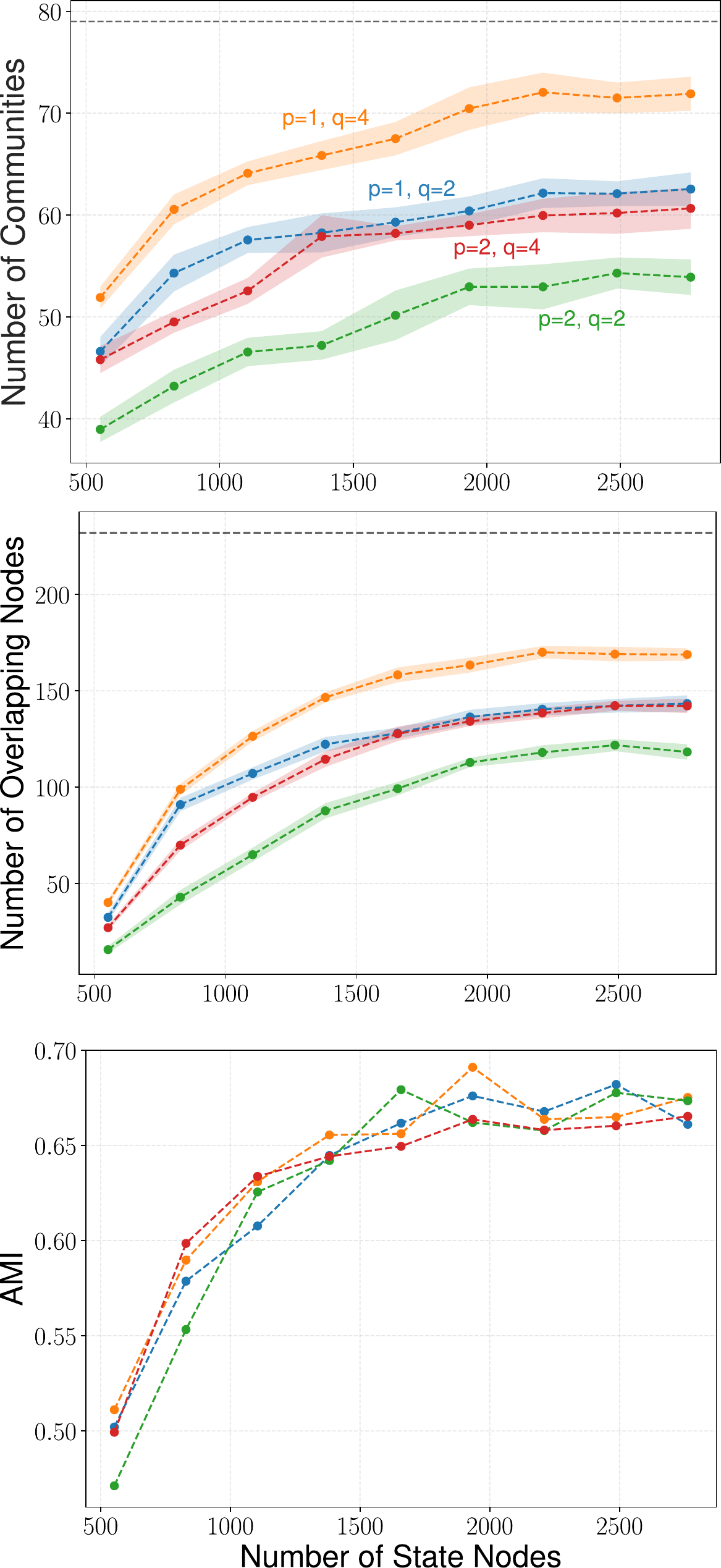}
    \caption{\textbf{Compact memory-biased models compared with a model using higher-order network data for taxi drives.} Four compact memory-biased models with varying $p$ and $q$ values for nine state-node budgets compared with a model using complete trigram data (dashed lines). The confidence intervals represent one standard deviation computed over 20 iterations.}
    \label{fig:taxi}
\end{figure}

As we increase the state-node budget, the compact model identifies more communities that overlap in more nodes (Fig.~\ref{fig:taxi}). The number of communities and number of overlapping nodes approach those found with complete trigram data. The AMI shows a similar trend: high-$q$ models, which best match the triangle-closing bias in the trigram data, achieve the highest AMI before levelling off.

The AMI plateaus below 1 for two reasons. First, we compare with one particular partition of the trigram data in a flat solution landscape. Second, uniform $p$ and $q$ parameters cannot capture the spatially varying memory in real taxi flows. While the compact model cannot fully replicate the spatial heterogeneity present in complete trigram data, it can capture dominant memory patterns and approximate the community structure.

\subsection{Model size}
Our aim is to recover the modular structure of the full second-order model using fewer state nodes. Adding more state nodes expands the space of possible partitions and typically yields higher compression but also longer search times. We use the codelength as a proxy for model quality to guide the selection of an appropriate model size.

\begin{figure*}[htp!]\centering\includegraphics[width=1.0\textwidth]{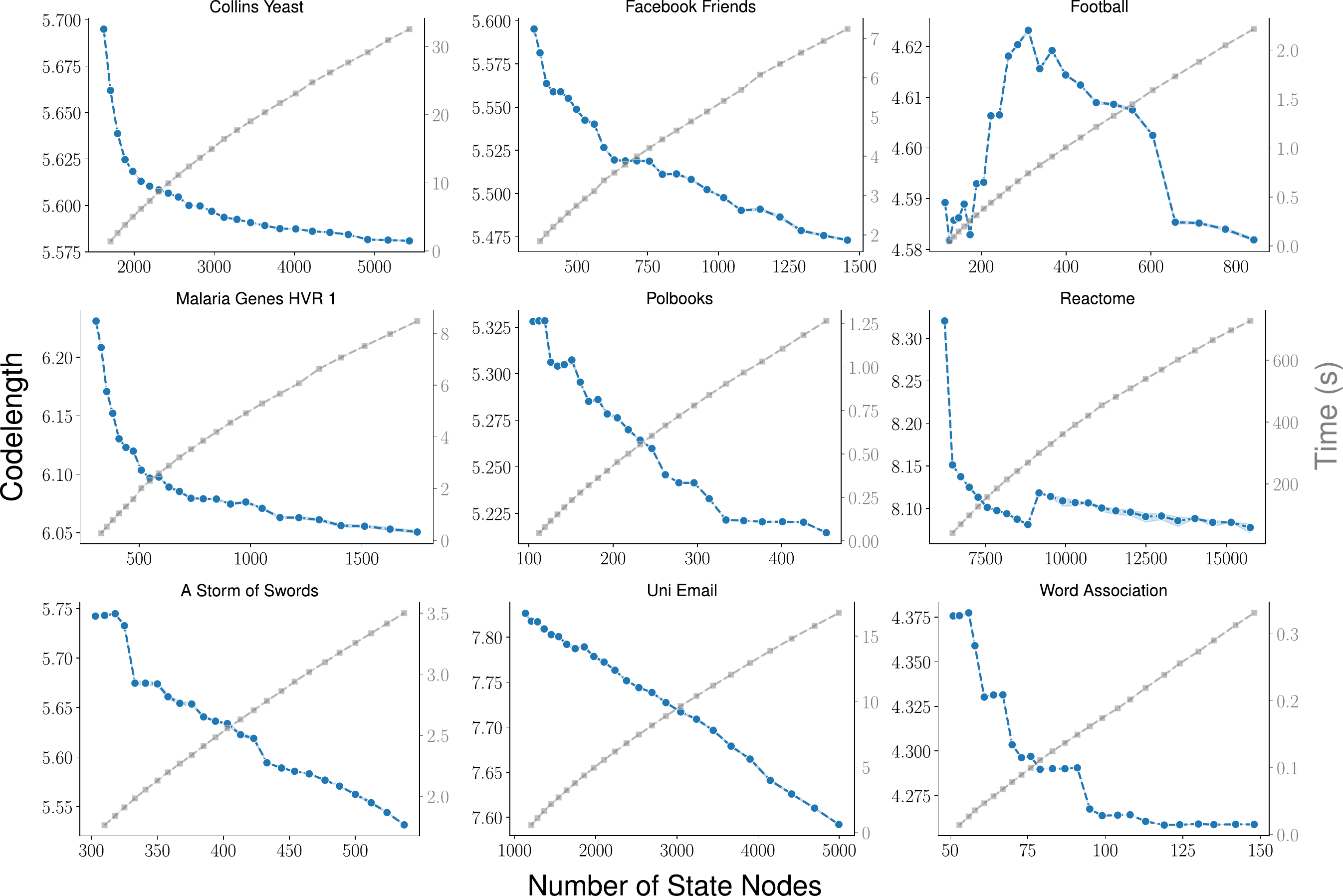}
    \caption{\textbf{Codelength and time to create the compact networks as a function of model size across nine real-world networks.} We used $p=1$ and $q=2$ for all networks except \textit{A Storm of Swords} and the football network where we used $q=4$. The confidence intervals represent one standard deviation computed over 20 iterations.}
    \label{fig:expanded_codelength}
\end{figure*}

To evaluate this trade-off in practice, we analyse nine different networks (\cref{fig:expanded_codelength}) and observe a consistent pattern. In most cases, the codelength drops sharply with the initial addition of state nodes, then gradually levels off as more are included. This drop suggests that substantial improvements can be achieved by moving beyond the simplest model $M_1$, but as the number of state nodes approaches that of $M_2$, the additional gains become marginal. Construct time increases sub-linearly with state-node budget, consistent with the divisive clustering algorithm's $\mathcal{O}(md)$ complexity. Infomap clustering then scales linearly with the number of links in the compact network~\cite{edler2017mapping}. These scaling properties define the practical trade-off: larger models achieve lower codelength but require longer construction and clustering, while compact models sacrifice some compression for faster analysis.

\section{Conclusion}
We introduce a method to detect overlapping communities in first-order networks by modelling memory-biased random walks. Drawing on principles from \textsl{node2vec} and higher-order network models, we avoid explicit random walk simulations by expressing memory constraints with state nodes. This representation enables ergodic network flow calculations and direct use of the community-detection algorithm Infomap for assigning state nodes to modules.
For a compact representation that balances sparsity and expressiveness, we lump state nodes based on an information loss criterion. Users can control this trade-off in two ways: by setting a state-node budget that limits the total number of nodes, or by specifying the acceptable level of information loss. By default, the method uses a compression threshold of 90 percent.

The bias parameters $p$ and $q$ in \textsl{node2vec} provide flexibility, allowing users to tailor the memory structure to the network at hand.
Applied to overlapping LFR and ABCD-o$^2$ benchmark networks, our method is robust across a wide range of parameter values. The AMI rises sharply as we add state nodes, demonstrating that 
compact models with about 50 percent of the full representation 
capture essential dynamics. Models successfully identify overlap 
when present, with the state-node requirement scaling naturally 
with the amount of overlap.
In real-world networks with overlapping community structures, we identify coherent and interpretable overlapping communities similar to those identified with complete higher-order network data.
While the uniform $p$ and $q$ parameters cannot capture node-specific memory patterns present in real higher-order data, our empirical results demonstrate that compact models with global parameters approximate community structures obtained from complete higher-order data.

\section{Acknowledgements}
We thank Jelena Smiljani{\'c}, Renaud Lambiotte and Tommy L{\"o}fstedt for helpful discussions and comments. M.L.\ was supported by the Wallenberg AI, Autonomous Systems and Software Program (WASP)\cite{wasp}, funded by the Knut and Alice Wallenberg Foundation. R.S. was funded by the Mathematical Institute at the University of Oxford. C.B.\ acknowledges funding from the Swiss National Science Foundation, grant 176938, and the German Federal Ministry of Education and Research, grant 100582863 (TissueNet). M.R. was supported by the Swedish Research Council under grant 2023-03705.

\section{Data Availability}
We collected the word association network from Ref.~\citenum{cfinder}, the \textit{A Storm of Swords} network from Ref.~\citenum{asoiaf}, the Amazon network from Ref.~\citenum{amazon}, the taxi drives from Ref.~\citenum{rosvall2014memory}, and downloaded the other networks from Ref.~\citenum{networks_skewed}.
The LFR and ABCD$+o^2$ benchmark source codes are available from Refs.~\citenum{lfr_code} and \citenum{abcd_code}.

\bibliographystyle{unsrt}

\onecolumngrid
\appendix
\renewcommand{\thefigure}{A\arabic{figure}}
\renewcommand{\thetable}{A\Roman{table}}
\setcounter{figure}{0}
\setcounter{table}{0}

\clearpage
\section*{Appendix}

\begin{table*}[htp!]
    \centering
    \caption{\textbf{Summary of the real-world networks used in the analysis. } }
    \begin{tabular}{lrrrr}
        \toprule
        \textbf{Networks} & Nodes & Edges & State Nodes (M$_2$) & State Nodes (M$_c$) \\
        \midrule
        collins\_yeast\cite{collins_yeast} & 1,622 & 9,070 & 18,140 & 5,436 \\
        facebook\_friends\cite{facebook_friends} & 362 & 1,988 & 3,976 & 1,457 \\
        football\_tsevans\cite{girvan2002community, evans2010clique} & 115 & 613 & 1,226 & 842 \\
        malaria\_genes/HVR\_1\cite{malaria_genes} & 307 & 2,812 & 5,624 & 1,748 \\
        polbooks\cite{polbooks} & 105 & 441 & 882 & 453 \\
        reactome\cite{reactome} & 6,327 & 147,547 & 292,320 & 15,745 \\
        storm\_of\_swords\cite{asoiaf} & 303 & 1,008 & 2,016 & 537 \\
        uni\_email\cite{uni_email} & 1,133 & 10,902 & 10,902 & 4,995 \\
        word\_association\cite{palla2005uncovering} & 51 & 149 & 298 & 148 \\
        \bottomrule
    \end{tabular}
    \label{tab:dataset_summary}
\end{table*}

\subsection{A storm of swords}
\label{appendix:storm_of_swords}
We list the 10 modules in the compact Storm of Swords network with at least 10 characters and name them based on the story arc they correspond to.
\begin{enumerate}
    \item \textbf{King's Landing (63 characters, 28.6\% flow)} Addam Marbrand, Aegon Targaryen (son of Rhaegar), Aemon Targaryen (Dragonknight), Aerys II Targaryen, Alayaya, Alerie Hightower, Amory Lorch, Anders Yronwood, Baelor I Targaryen, Balon Swann, Boros Blount, Brella, Bronn, Butterbumps, Cersei Lannister, Chataya, Doran Martell, Eddard Stark, Elia Martell, Ellaria Sand, Galyeon of Cuy, Garlan Tyrell, Gregor Clegane, High Septon (Tyrions), Jaime Lannister, Joanna Lannister, Joffrey Baratheon, Kella, Kevan Lannister, Lancel Lannister, Leonette Fossoway, Lollys Stokeworth, Loras Tyrell, Lysa Arryn, Mace Tyrell, Maegor I Targaryen, Mandon Moore, Margaery Tyrell, Mathis Rowan, Meryn Trant, Mordane, Myrcella Baratheon, Oberyn Martell, Olenna Redwyne, Osmund Kettleblack, Paxter Redwyne, Podrick Payne, Pycelle, Renly Baratheon, Robert Baratheon, Sansa Stark, Shae, Stannis Baratheon, Symon Silver Tongue, Tanda Stokeworth, Tommen Baratheon, Tyrion Lannister, Tysha, Tywin Lannister, Urswyck, Varys, Viserys Targaryen, Willas Tyrell
    
    \item \textbf{The Wall (35 characters, 13.0\% flow)} Aemon Targaryen (Maester Aemon), Alliser Thorne, Bannen, Bedwyck, Blane, Bowen Marsh, Byam Flint, Chett, Clubfoot Karl, Clydas, Cotter Pyke, Craster, Denys Mallister, Dirk, Donal Noye, Dywen, Eddison Tollett, Gilly, Grenn, Hobb, Janos Slynt, Jeor Mormont, Jon Snow, Lark, Mance Rayder, Othell Yarwyck, Pypar, Qhorin Halfhand, Ragwyle, Samwell Tarly, Small Paul, Softfoot, Stannis Baratheon, Watt, Wynton Stout

    \item \textbf{The Riverlands (42 characters, 12.3\% flow)} Aegon Frey (son of Stevron), Arya Stark, Balon Greyjoy, Brenett, Brynden Tully, Catelyn Stark, Dacey Mormont, Desmond Grell, Edmure Tully, Elmar Frey, Gawen Westerling, Hoster Tully, Jaime Lannister, Jeyne Westerling, Jon Umber (Greatjon), Jon Umber (Smalljon), Lothar Frey, Maege Mormont, Marq Piper, Olyvar Frey, Perwyn Frey, Petyr Frey, Ramsay Snow, Raynald Westerling, Rickard Karstark, Robb Stark, Robin Ryger, Rodrik Cassel, Rollam Westerling, Rolph Spicer, Roose Bolton, Roslin Frey, Ryman Frey, Stevron Frey, Sybell Spicer, Tansy, Tristifer IV Mudd, Tywin Lannister, Utherydes Wayn, Vyman, Walder Frey, Walder Rivers

    \item \textbf{Arya's adventures (12 characters, 7.4\% flow)} Arya Stark, Cersei Lannister, Dunsen, Gregor Clegane, Ilyn Payne, Joffrey Baratheon, Meryn Trant, Mycah, Polliver, Rafford, Sandor Clegane, Tickler

    \item \textbf{North of the Wall (23 characters, 6.2\% flow)} Arson, Big Boil, Dalla, Errok, Gendel, Gorne, Grigg, Harma, Jarl, Jon Snow, Joramun, Mance Rayder, Munda, Orell, Rattleshirt, Ryk, Styr, Tormund, Val, Varamyr, Weeper, Ygritte, Zei

    \item \textbf{Essos (25 characters, 5.9\% flow)} Arstan, Barristan Selmy, Belwas, Ben Plumm, Cleon, Daario Naharis, Daenerys Targaryen, Drogo, Ghael, Grazdan mo Eraz, Grey Worm, Groleo, Illyrio Mopatis, Irri, Jhiqui, Jorah Mormont, Kraznys mo Nakloz, Mero, Missandei, Oznak zo Pahl, Rhaegar Targaryen, Rhaegel Targaryen, Rhaella Targaryen, Robert Baratheon, Viserys Targaryen

    \item \textbf{The Brotherhood without Banners (20 characters, 5.7\% flow)} Anguy, Arya Stark, Beric Dondarrion, Gendry, Greenbeard, Harwin, Hot Pie, Husband, Jack Be Lucky, Jack Bulwer, Lem, Mad Huntsman, Merrit, Pate (Old), Ravella Swann, Sandor Clegane, Sharna, Tansy, Thoros of Myr, Tom of Sevenstreams

    \item \textbf{Dragonstone (19 characters, 5.0\% flow)} Alester Florent, Andrew Estermont, Axell Florent, Azor Ahai, Cressen, Davos Seaworth, Devan Seaworth, Edric Storm, Gerald Gower, Guncer Sunglass, Khorane Sathmantes, Lamprey, Melisandre, Porridge, Pylos, Salladhor Saan, Selyse Florent, Shireen Baratheon, Stannis Baratheon

    \item \textbf{Jon's Watch (10 characters, 3.3\% flow)} Dick Follard, Grenn, Hareth (Moles Town), Jon Snow, Kegs, Mully, Owen, Pypar, Satin, Spare Boot

    \item \textbf{Winterfell (11 characters, 2.7\% flow)} Arya Stark, Benjen Stark, Bran Stark, Brandon Stark, Eddard Stark, Jon Snow, Luwin, Nan, Rickon Stark, Robb Stark, Theon Greyjoy

\end{enumerate}

\begin{table}
    \centering
    \caption{\textbf{Central characters in A Storm of Swords.} The characters with more than 2 percent flow each and the modules they belong to.}
    \begin{tabular}{lcl}
        \toprule
        Character & Total flow & Module (fraction of flow) \\ 
        \midrule
        Tyrion Lannister & 0.035 & King's Landing (1.0)\\
        \midrule
        Joffrey Baratheon & 0.030 &  \makecell[l]{King's Landing (0.82) \\ Arya's adventures (0.18)}\\ \midrule
        Jon Snow & 0.030 & \makecell[l]{North of the Wall (0.42)\\ The Wall (0.26)\\ Jon's Watch (0.19)\\ Winterfell (0.13)}\\ \midrule
        Sansa Stark & 0.027 & King's Landing (1.0) \\ \midrule
        Robb Stark & 0.027 & \makecell[l]{The Riverlands (0.90)\\ Winterfell (0.10)}\\ \midrule
        Jamie Lannister & 0.025 & \makecell[l]{King's Landing (0.74)\\ The Riverlands (0.15)\\ Module \# 11 (0.11) }\\ \midrule
        Cersei Lannister & 0.023 & \makecell[l]{King's Landing (0.73)\\Arya's adventures (0.27)}\\ \midrule
        Arya Stark & 0.022 & \makecell[l]{Arya's adventure (0.42) \\The Brotherhood without Banners (0.31)\\The Riverlands (0.13)\\Winterfell (0.13)}\\ \midrule
        Catelyn Stark & 0.020 & The Riverlands (1.0)  \\
        \bottomrule
    \end{tabular}
    \label{tab:storm_of_swords_top15}
\end{table}

\end{document}